\documentclass[a4paper]{jpconf}
\usepackage{graphicx}
\begin{document}
\title{Baryon charge asymmetry at LHC: String Junction transfer in proton reactions and SJ torus as DM candidate}

\author{Olga I. Piskounova}

\address{P.N. Lebedev Physics Institute, Moscow 119991, Russia}

\ead{piskoun@sci.lebedev.ru}

\begin{abstract}
The asymmetry of baryon/antibaryon production has been measured in many proton-proton, pion-proton and electron-proton experiments. In the framework of Quark-Gluon String Model (QGSM) the energy dependence of asymmetry tells us about the value of $\alpha_{SJ}$(0), the intercept of String Junction Regge trajectory. In previous QGSM study, the value of intercept has been estimated as 0.5 < $\alpha_{SJ}$(0) < 0.9. Here, SJ behaviors are accumulated in the model based on topological expansion in order to build a neutral object with zero baryon charge. By the way, QCD mass falling under the event horizon of Black Hole (BH) should be symmetric, or in other words, have no charge information. The baryon junctions are easily combinable with antibaryon ones in hexagons. Topologically, hexagon net can coherently cover only the torus surface. The net on the torus has discrete number of baryon/antibaryon junctions. This is only parameter that marks the mass/energy level of this object. It looks like DM particle, is not it? In high energy collisions at LHC, such pomeron loops are to be produced approximately in 1.2 percent of inelastic events. Furthermore, the torus configurations of matter have been revealed in many bright events in space. As an example, Chandra experiment has detected such dense "doughnut" near the event horizon of Super Massive Black Hole (SMBH), which X-ray radiation is screened on 40 percent's. This topological symmetry model of DM seems rather realistic and can help us to deal with an "arm wrestling" between the stiffness of toroid structure of QCD matter and the pressure of gravitational singularity at extremely heavy masses. On the other hand, the instabilities in structure of matter in SMBHs can cause the bursts of giant relativistic hadron jets with the masses of order the own BH mass.
\end{abstract}
\section{Introduction}

Quark-Gluon String Model allows us to calculate the energy spectra of baryons produced in various hadron-hadron collisions. The formulas of QGSM are based of topological expansion of QCD diagrams \cite{veneziano} and Regge asymptotics of fragmentation of quark-gluon strings that gives the hadrons of all sorts. The Model accounts the quark composition of interacting particle, which influenses on the asymmetries between hadron spectra.
As it was observed in many experiments, there is nonzero asymmetry of baryon/antibaryon spectra left in central rapidity region at proton-proton collisions of high energies. This asymmetry can not be explaned above the energy 200 GeV as the migration of proton diquark to low rapidity area. We invented in the previous papers \cite{kaidalov, baryonasymmetry} that positive baryon charge can be transfered just with String Junction (SJ) from proton projectiles. The previous QGSM calculations have described the long-serviving positive baryon asymmetry at LHC energies.
In this paper we intend to study another manifestation of SJ behaviors in hadron collisions and to show what can these objects add into the QCD processes on the galaxy scales. 

\section{String Junction and the Positive Baryon Production Asymmetry at LHC}

\begin{figure}[h]
\centering
  \includegraphics[width=14pc]{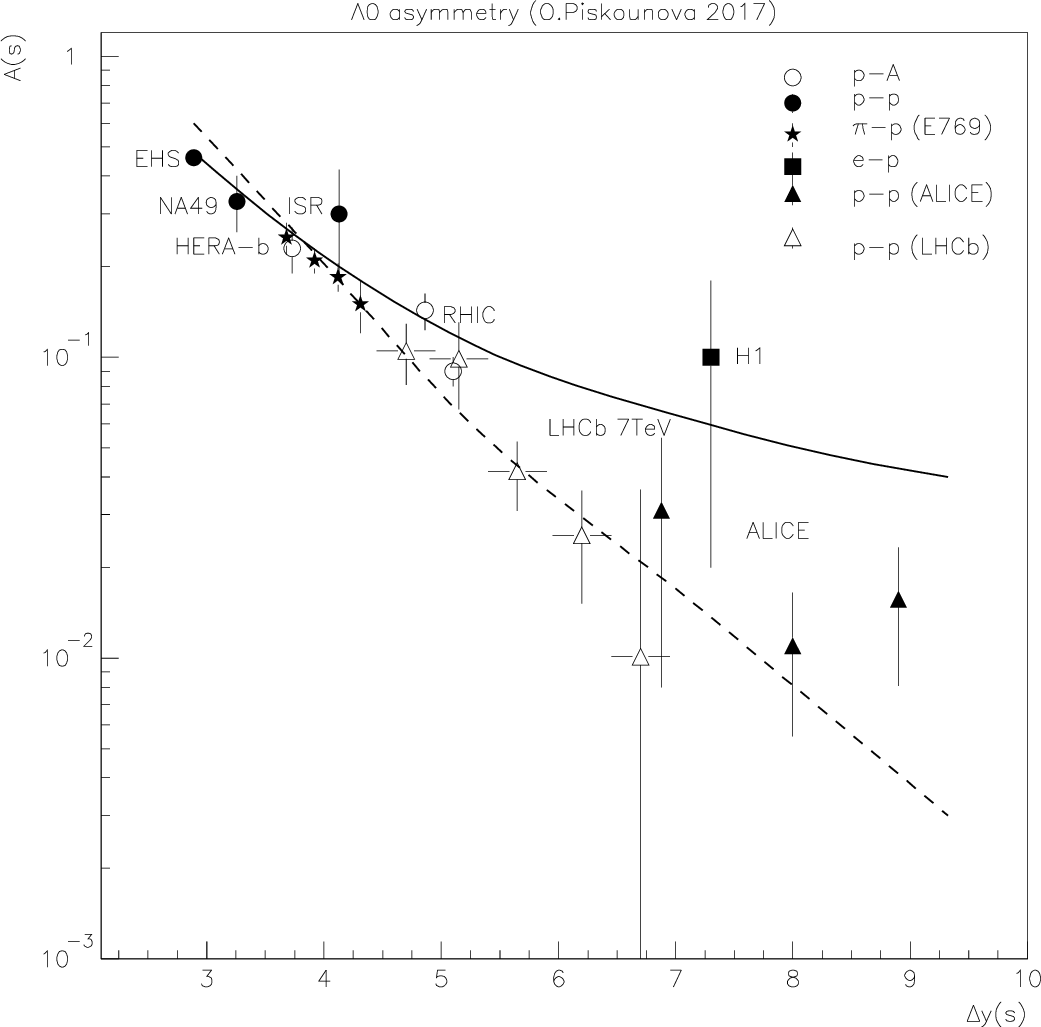}
  \caption{Hyperon asymmetry dependence on the the rapidity distance from center to the beam proton. Two theoretical lines correspond to the values of $\alpha_{SJ}$: 0.9 - solid line and 0.5 - dashed line.}
  \label{lamasymmetry}
\end{figure}

The connection of three gluons of "Mercedes" type, or String Junction (SJ), seems playing the important role for the multi particle production in our positive-baryon-charge world as well as in collider experiments. First of all, SJ brings the positive baryon charge and generates the valuable asymmetry between baryon and antibaryon spectra at the central rapidity area in any collisions with proton projectile. This asymmetry is surviving even at LHC energies that was confirmed by available collider data, see figure~\ref{lamasymmetry}. The asymmetry data of ALICE have been added recently \cite{alice}. 

The possibility to bring proton baryon charge with the diquark 
has disappeared at the energies above $\sqrt{s}$ = 200 GeV. Such a way, the parameter $\alpha_{SJ}$ plays the main role in the fragmentation of SJ that cause  the extra baryons at the central rapidities on LHC.

\section{Topological Expansion and Pomeron Exchange}

\begin{figure}
\centering
\includegraphics[width=8pc]{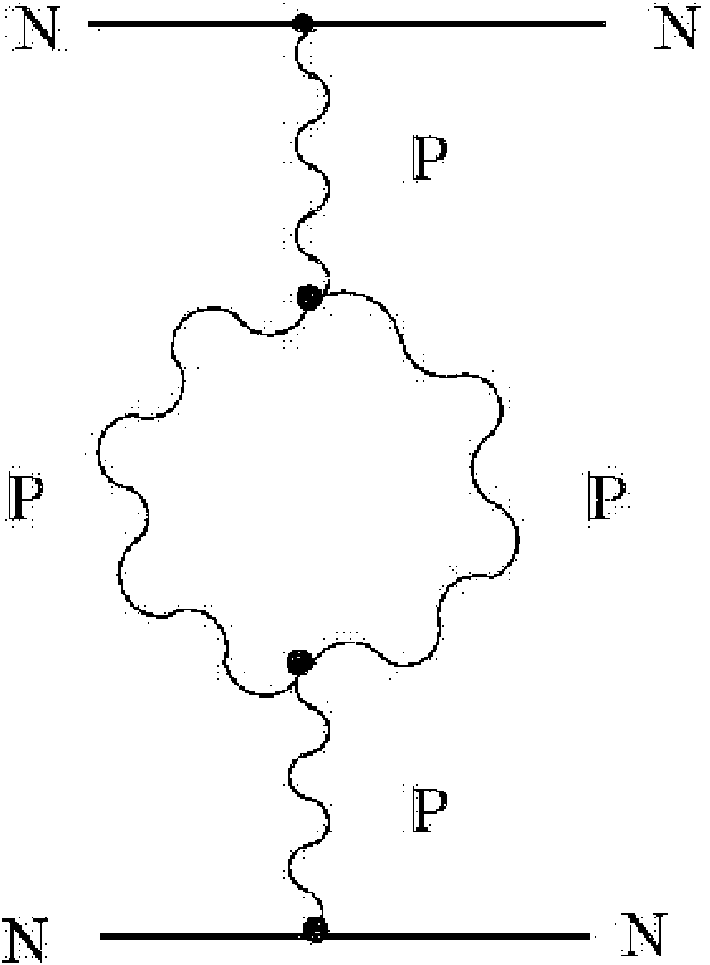}
  \caption{Pomeron loop in the center of one-pomeron exchange.}
  \label{pomeronloop}
\end{figure}

Pomeron exchange can be shown as topological QCD diagram and is responsible for multi particle production in p-p collisions at LHC energies. It was drawn in the topological expansion \cite{topologyexp} as the cylindrical net of gluon exchanges with the random amount of quark-antiquark loops also inserted into the net. The topological expansion gives the chance to classify the contributions from general diagrams of multi particle production in the hadron interactions. The developing of this expansion has practically allowed us to build Quark-Gluon String Model (QGSM). First orders of topological expansion were presented in \cite{veneziano}, where the third order is named pomeron with handle The case of double diffraction dissociation in this presentation looks like the cylinder of one pomeron exchange with the handle that is topologically similar to pomeron torus, see figure~\ref{pomeronloop}. Due to 1/Nc expansion, Pomeron exchange used to take 1/9 from the leading contribution of planar diagram that is first order of the expansion of quark-gluon diagrams. But the quark annihilation diagram is dying out with energy as $s^{-0.5}$ due to the Regge behavior of processes with quark annihilation. In the QCD phenomenology the cross sections of one pomeron exchange is to be growing as $s^{\Delta_P}$, where the pomeron trajectory parameter $\Delta_P = \alpha_P(0)-1$ = 0.12. Such a way the second order diagrams with the pomeron exchange will dominate at high energies.

\section{ Double Diffractive Dissociation as an Exchange with Pomeron Torus}

Double diffraction dissociation (DD) is a next order in the topological expansion after the pomeron exchange and should be presented as one pomeron exchange with the pomeron loop (PL) in the center, see figure~\ref{pomeronloop}. Actually, this construction is the topological cylinder with a handle that takes $(1/9)^2$ from total pomeron exchange cross section or 1,2 percent of $\sigma_{prod}$) at LHC energies. This interesting object should be considered separately in order to reveal some remarkable features for the experimental detection.
Preliminary, ATLAS experiment has measured the distribution of 
rapidity gaps in diffraction processes, see \cite{myatlastalk}. The fluctuations at the big gaps can be explaned as the existing the energy levels of PL, see cite{atlas}. More detailed resarch is needed. 

If the central PL is not cut, we are having the DD spectra of produced hadrons: two intervals at the ends of rapidity range, which are populated with multi particle production, and the valuable gap in the center of rapidity, see figure~\ref{multiplicity_with_gap}.
Otherwise, we have the same gap in the center that is populated with doubled multiplisity, if both sides of PL are cut. 
 
\begin{figure}[h]
\centering
\includegraphics[width=14pc]{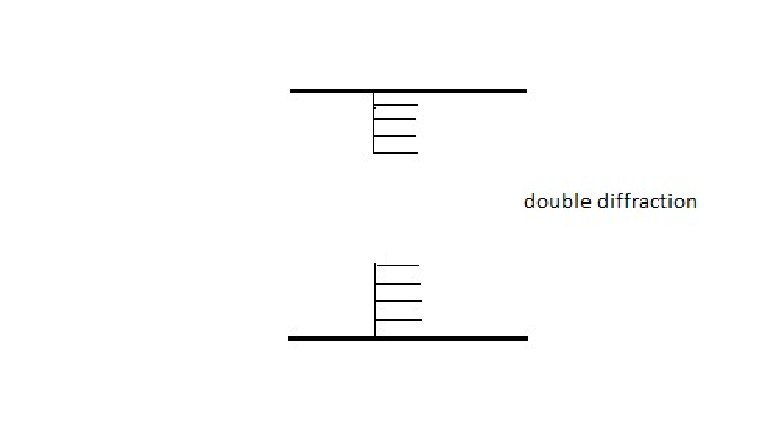}
\caption{\label{multiplicitywithgap}The diagram of particle production in DD with the rapidity gap.}
\end{figure}
   
In the other two cases with one cut side of PL we are getting multi particle production with the ordinary multiplicity.

\section{Junction-Antijunction Hexagon Net and Discrete Dynamics of SJ torus}

\begin{figure}[h]
\centering
\includegraphics[width=14pc]{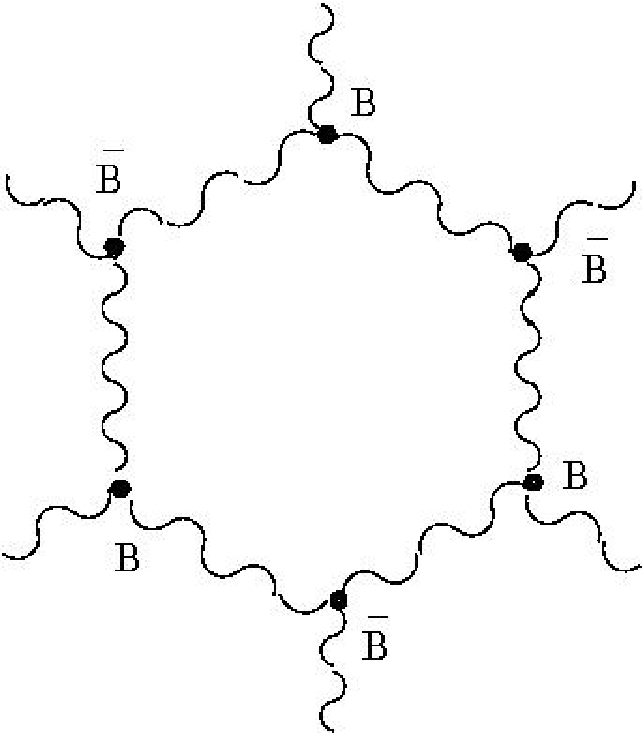}
\caption{\label{onecell}One cell in the hexagon net with SJs and antiSJs.}
\end{figure}

As we remember, the pomeron cylinder is built by  the net of gluon exchanges, see figure~ref{myPhD}. Let us consider only three gluon connections on the surface of PL torus. This SJ type of gluon vertices has been studied in the early research \cite{baryontorus}. Since this object brings the baryon charge, the antiSJ also exists and brings the charge of antibaryon. The only charge-neutral way to construct the net from SJ's and antiSJ's is hexagon where antibaryon charge is following the baryon one as it is shown in the figure~\ref{onecell}.
 
It seems very entertaining game to close certain number of hexagons on the surface of torus! The net of six hexagons can cover the torus, as it is shown in the
figure~ref{torus}. 

\begin{figure}[h]
\centering
\includegraphics[width=14pc]{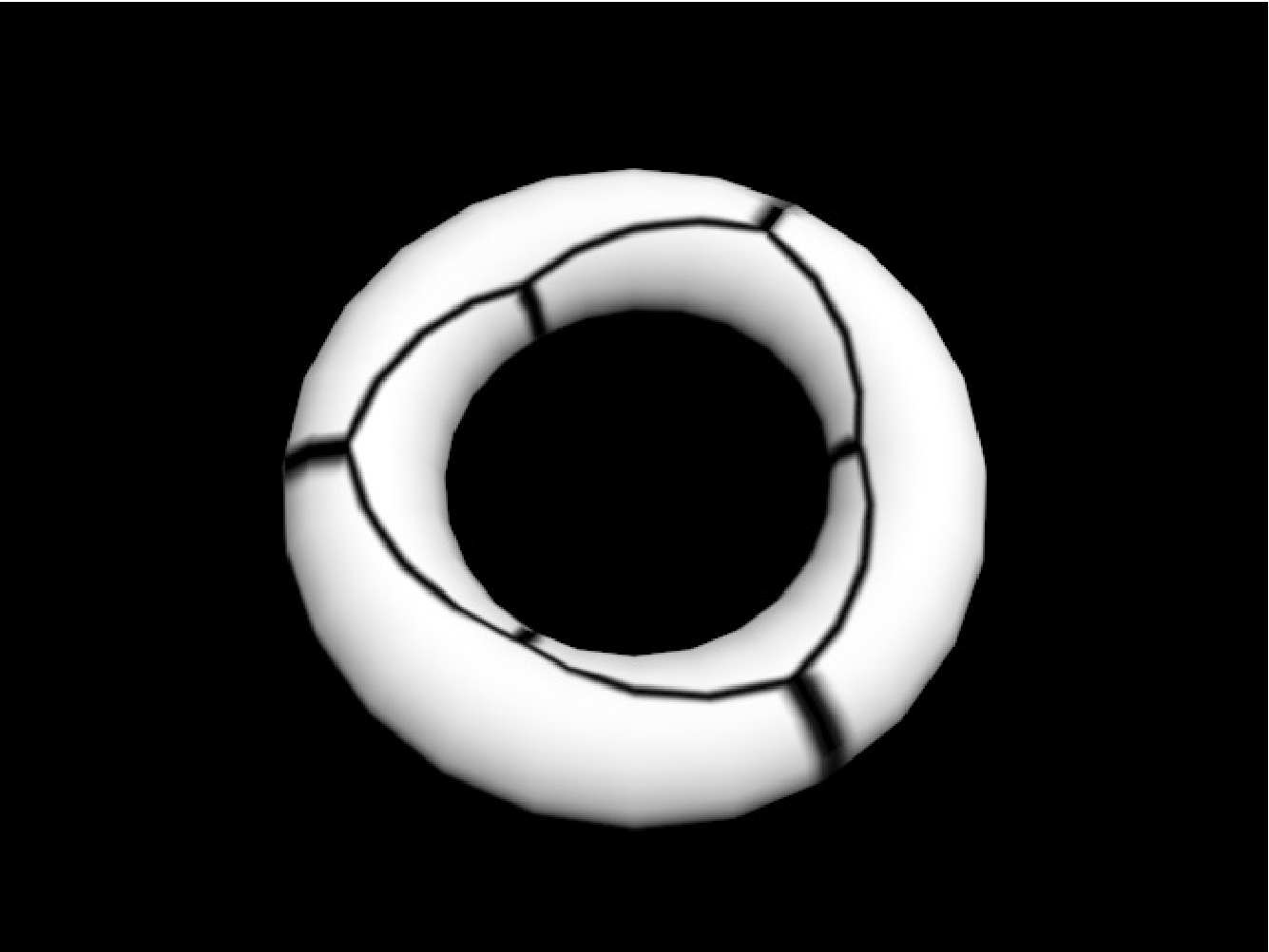}
\caption{\label{torus}The net of 6 hexagons on the surface of pomeron torus.}
\end{figure}

If people are trying to match the eligible number of hexagons, it becomes clear that there is a row with discrete numbers of hexagons: Hexnum = 4, 6, 12, 16, 20, 30, 36, 42, 56 and so forth. Actually, so many possibilities mean the easy way to add 
few hexagons and build the next torus. Such a way, the SJ torus has certain levels of energy.
 
\section{SJ torus as DM candidate}

The are few reasons why we can take SJ torus as a candidate for dark matter:
1) it has certain states of energy (or mass);
2) it has nor baryon nor other charge;
3) it may be rather compact and decays rarely or iven does not decay, if of great mass;
4) if it is very small and dense, this torus should easy penetrate through atoms and nucleus of ordinary matter.
May be the lightest SJ toruses (SJ+antiSJ) are just absorbed by protons. The heavy ones are concentrated in the center of neutron stars under the experemely high density. Futhermore, the best possibility for this sort of matter is to surround the Black Holes. What is especially remarkable, this dark matter "particle" consists of ordinary baryon matter in the same way as a diamond consists of carbon!

\begin{figure}[t]
\centering
\includegraphics[width=14pc]{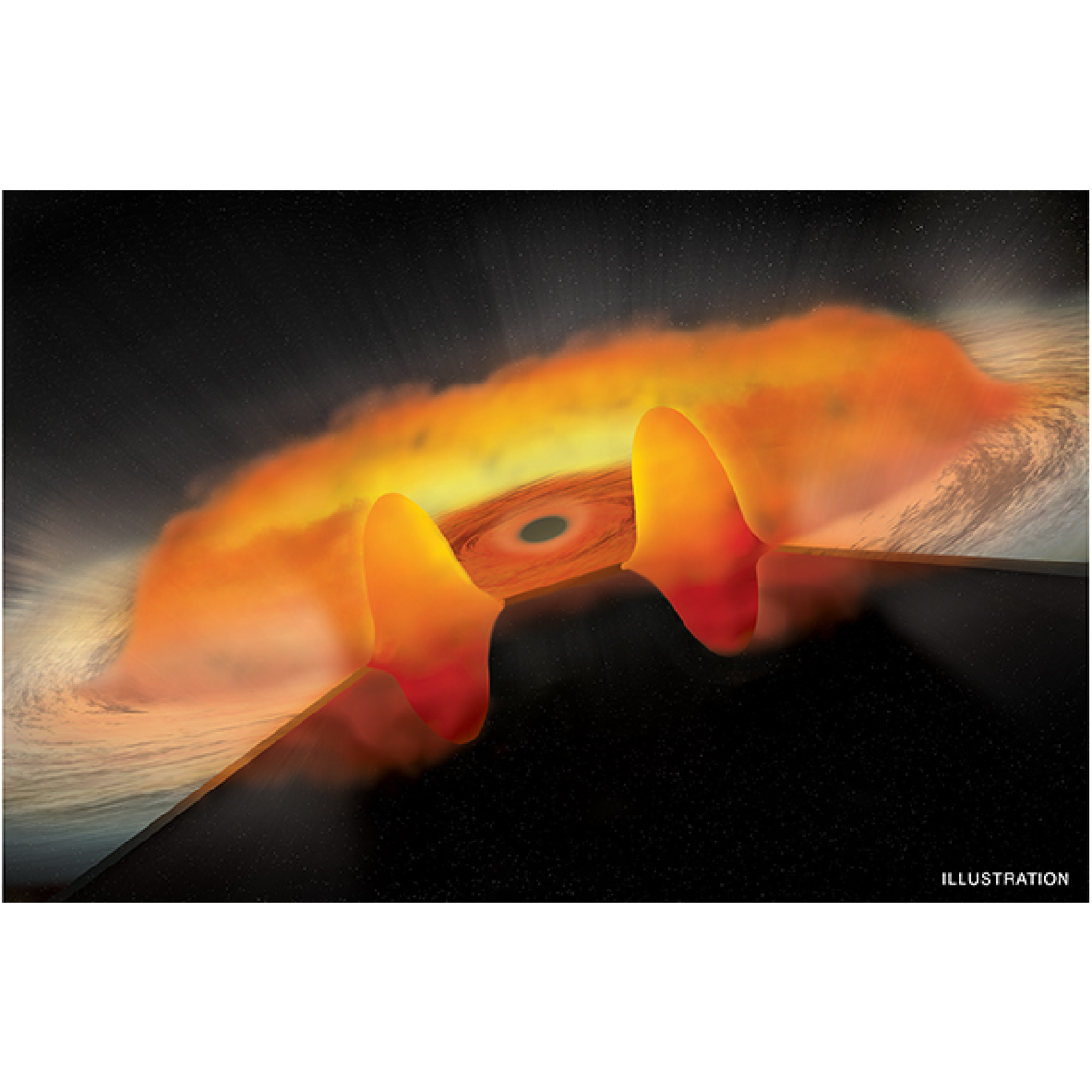}
\caption{\label{chandra}Illustration of torus configuration of matter that has revealed by Chandra experiment.}
\end{figure}

\section{Manifestations of Toroidaly Organized Matter in Space Observations}

After the above discours it seems very curious to find the toroidaly shaped matter in the high energy events in astrophysics. There are few observations near the Massive Black Holes (SMBH). As an example, here are two of them: a) at the horizon of SMBH, see figure~\ref{chandra} \cite{chandra} and b) inside the relativistic jet radiated from SMBH \cite{ngc6109}. The last gives a key to the understanding how the relativistic jets are created.
As I remember, the star formation from the remnants of Supernova 1987 goes also withing the toroid shape. 

\begin{figure}{h}
\centering
\includegraphics[width=14pc]{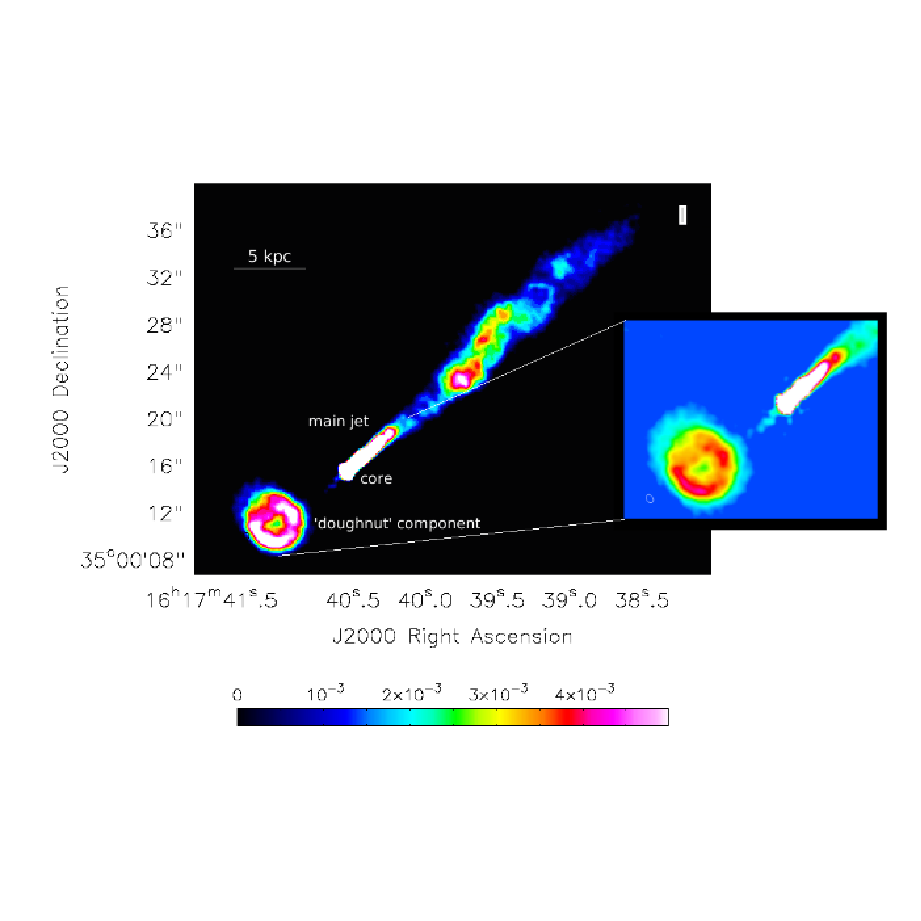}
\caption{\label{ngc6109}Torus structure of jet from radio galaxy
NGC6109 is a recent observation of baryon matter at the extremal conditions.}
\end{figure}

Finally, there are few words on the possible dynamics of SJ torus inside BH. Since the toroidally organized DM is dense but elastic construction, it can be squashed by strong gravity and return to the lower level of energy that is followed by the radiation of more than one third of its mass into the giant jets with the torus configurations, see figure~\ref{ngc6109}.
 
\section{Conclusions}

Concluding about the behaviors of SJ, we should emphasize on:
1) String Junctions bring baryon charge in the high energy proton-proton collisions and cause the positive baryon asymmetry at LHC energies;
2) SJ can be organized with anti SJ in the neutral structures
  (hexagon nets) that are able to build SJ torus with zero baryon charge;
3) this toroidally-organized QCD matter behaves as DM: SJ torus has discrete levels of energy (mass), it is very compact, penetrative through the matter and rarely decays into the structures of lower mass; 
5) giant torus structures have been observed near SMBHs;
6) the presented topological-symmetry model can explain as well an "arm wrestling" of BH gravity and QCD matter at the extremely high masses and densities;
7) this dynamical interaction produces the relativistic jets directly from the center of SMBHs.

\section {Acknowledgments}

Author expresses her gratitude to Prof. Michael Polikarpov with
his students and colleagues, whose enthusiasm in graphen studies
gave me the idea of "compactificated" quark-gluon (pomeron) string.

\end{document}